\documentclass[pra,onecolumn,notitlepage,nofootinbib]{revtex4-1}

\usepackage[utf8]{inputenc}
\usepackage[english]{babel}
\usepackage[T1]{fontenc}

\usepackage{graphicx}
\usepackage{amssymb,amsfonts,amsmath,enumerate,amsthm}

\usepackage{lmodern}
\usepackage{braket}

\newcommand{\ketbra}[2]{\ket{#1}\bra{#2}}

\def\A{\mathcal{A}}
\def\B{\mathcal{B}}
\def\C{\mathcal{C}}
\def\E{\mathcal{E}}
\def\H{\mathcal{H}}
\def\M{\mathcal{M}}
\def\O{\mathcal{O}}
\def\S{\mathcal{S}}
\def\W{\mathcal{W}}
\def\Cset{\mathbb{C}}
\def\Nset{\mathbb{N}}
\def\Rset{\mathbb{R}}
\def\Tr{\operatorname{Tr}}
\def\Salicru{H_{(h,\phi)}}
\def\SalicruQ{\mathbf{H}_{(h,\phi)}}

\newtheorem{definition}{Definition}
\newtheorem{proposition}{Proposition}

\begin{document}

\title{Generalized entropies in quantum and classical statistical theories}

\author{M. Portesi\textsuperscript{1}, F. Holik\textsuperscript{1}, P.W. Lamberti\textsuperscript{2}, G.M. Bosyk\textsuperscript{1}, G. Bellomo\textsuperscript{3} and S. Zozor\textsuperscript{4}}

\affiliation{
\textsuperscript{1}Instituto de Física La Plata, UNLP, CONICET, Facultad de Ciencias Exactas, 1900 La Plata, Argentina}

\affiliation{\textsuperscript{2}Facultad de Matem\'{a}tica, Astronom\'{i}a, F\'{i}sica y
Computaci\'{o}n (FAMAF), UNC, CONICET, C\'{o}rdoba, Argentina}

\affiliation{\textsuperscript{3}CONICET-Universidad de Buenos Aires, Instituto de Investigación en Ciencias de la Computación (ICC), Buenos Aires, Argentina}

\affiliation{\textsuperscript{4}Laboratoire Grenoblois d'Image, Parole, Signal et Automatique
(GIPSA-Lab), CNRS, Saint Martin d'H\`eres, France}

\begin{abstract}
We study a version of the generalized $(h,\phi)$-entropies, introduced by Salicr\'{u} \textit{et al}, for a wide family of probabilistic models that includes quantum and classical statistical theories  as particular cases. We extend previous works by exploring how to define $(h,\phi)$-entropies in infinite dimensional models.
\end{abstract}

\date{\today}

\maketitle

\section{Introduction}\label{intro}

The  concept of  entropy has  been widely  used in the physics
literature. But  it has also been  applied in information  theory.
An important example of  a recent development combining  both fields
of research  is given by quantum information theory \cite{NieChu10}.
In this field, quantum versions of information  measures play  a key
role. The quantum counterpart of Shannon entropic measure is the von
Neumann entropy~\cite{vNeu27,Holik-Plastino-Saenz-VNF,Watanabe}.
Also other measures have been adapted to the quantum realm in
different
contexts~\cite{Ren61,Tsa88,CanRos02,HuYe06,Kan02,SR-Paper}. Entropic
measures are important in several fields of research. They find
applications in the study of:
\begin{itemize}
\item  uncertainty  measures (as  is  the case  in  the study  of
  uncertainty relations~\cite{Uff90,ZozBos13,ZozBos14}),
\item         different        formulations         of         the        MaxEnt
  principle~\cite{Jaynes-Book,HeinQL,QuantalEffectsandMaxEnt,HolikIJGMMP},
\item                 entanglement                 measuring                 and
  detection~\cite{HorHor94,AbeRaj01:01,TsaLlo01,RosCas03,BenZyc06,Hua13,OurHam15},
\item  measures        of        mutual
  information~\cite{Yeu97,ZhaYeu98,Car13,GroWal13,Watanabe2}
\item    the    theory   of    quantum    coding    and   quantum    information
  transmission~\cite{SR-Paper,AhlLob01,Watanabe}.
\end{itemize}

In the theory of classical information measures, Salicr\'u
$(h,\phi)$-entropies~\cite{SalMen93} are, up to now, the most
generalized extension containing the Shannon~\cite{Sha48},
R\'enyi~\cite{Ren61} and Tsallis~\cite{Tsa88} entropies as
particular examples.

A finite dimensional quantum version of the $(h,\phi)$-entropies was
advanced and thoroughly studied in~\cite{QINP-paper}. A
generalization of the $(h,\phi)$-entropies to arbitrary finite
dimensional probabilistic models was introduced
in~\cite{Entropy-PaperGeneralized}
(see~\cite{Ohya-1989,Entropy-generalized-II,HeinQL} for
generalizations of more restricted families of entropic measures to
different frameworks).  In this short paper we extend the previous
definitions of $(h,\phi)$-entropies so as to include infinite
dimensional models.

The  paper is  organized  as follows. In
Section~\ref{ProbabilisticModels}  we introduce preliminary  notions
of  generalized   probabilistic models  and decomposition  theory.
In Section~\ref{s:Review}  we  discuss  the  classical formulation
of  $(h,\phi)$-entropies  and   provide  a definition  of  quantum
$(h,\phi)$-entropies   that includes    infinite   dimensional
models.    In Section~\ref{s:GeneralizedEntropies},   we   define
$(h,\phi)$-entropies   for probabilistic  theories whose states form
compact  convex  sets. Section~\ref{s:Conclusions} is devoted to
some concluding remarks.

\section{Probabilistic models}\label{ProbabilisticModels}

The description of quantum mechanical systems makes use of a family
of probabilistic models that can be radically different from those
originated  in classical statistical theories. It  is easy to show
that  both quantum and classical  state spaces are convex
sets~\cite{Entropy-PaperGeneralized}. Indeed, this result is much
more general: in the approach to physical theories based  on von
Neumann
algebras~\cite{HalvorsonARQFT,RedeiSummersQuantumProbability} the
sets of states are also convex.  The canonical example of a von
Neumann algebra is given by the set $\B(\H)$ of bounded operators
acting on a separable Hilbert space $\H$.  Due to von  Neumann
double commutant
theorem\cite{RedeiSummersQuantumProbability}\footnote{Given a
subset $M \subseteq
  \B(\H)$, the commutant of $M$ is defined as  $M' = \{A \in \B(\H) \: | \: AB -
  BA = 0 , \:\forall \, B  \in M\}$.}, it is possible to define a von
Neumann algebra as a $\ast$-subalgebra\footnote{For bounded
operators the $\ast$
  operation means just taking the adjoint  of a given operator (i.e., $A^\ast :=
  A^\dag$).  Thus,  the condition ``$\ast$-subalgebra'' reads  ``is a subalgebra
  that  is  closed  under  the  adjoint  operation''.}   $\W  \subseteq  \B(\H)$
satisfying $\W'' = \W$~\cite{Yngvason2005-TypeIIIFactors}.  $\B(\H)$
is not the  only  example of  a  von  Neumann algebra.   By
appealing  to a  dimension function, irreducible von Neumann
algebras can be classified in terms of factors of Type I, II  and
III~\cite{HalvorsonARQFT}. Only Type  I factors  appear in standard
quantum mechanics: the set of matrices of a complex finite
dimensional Hilbert space and  $\B(\H)$ (in the infinite dimensional
case), are examples of Type I factors. But other factors may appear
in the study of models of quantum mechanics involving infinitely
many  degrees of  freedom  (as is the case in quantum   field
theory~\cite{HalvorsonARQFT,Yngvason2005-TypeIIIFactors} and quantum
statistical  mechanics~\cite{Bratteli}).   A commutative  von
Neumann algebra can be  used to describe the algebra of  observables
of a classical probabilistic theory.  States in general von Neumann
algebras are defined in the standard way: a  state $\nu: \W
\longrightarrow \Cset$  is a continuous positive linear functional
such that  $\nu(\mathbf{I})=1$, with $\mathbf{I}$ the identity
operator over $\W$. Positivity means  that $\nu\left(A^\ast A\right)
\geq 0$ for all $A \in \W$.

All  von Neumann  algebras are  particular examples  of
C$^\ast$-algebras (see, e.g.,~\cite{Bratteli}). A
\emph{C$^\ast$-algebra} $\M$ is defined as a complex Banach algebra
endowed with an $\ast$ involution satisfying $(\alpha a + \beta
b)^\ast = \bar{\alpha} a^\ast + \bar{\beta} b^\ast$, $(ab)^\ast =
b^\ast a^\ast$ and $\|a a^\ast\|  = \|a\| \|a^\ast\|$, for all $a,b
\in  \M$ and $\alpha, \beta \in \Cset$.   All C$^\ast$-algebras can
be represented  as $\ast$-subalgebras of $\B(\H)$, closed under the
norm operator  topology. It is possible to show that, if the algebra
$\M$ is unital, then,  the set of states $\C(\M)$  is convex and
compact  (in the weak$^\ast$  topology \cite[Chap.2]{Bratteli}).
Furthermore,  due to the Krein-Milman theorem \cite[Chap.1]{Phelps},
the state space of a unital C$^\ast$-algebra $\M$  is the
weak$^\ast$ convex hull of its extreme points $\E(\C(\M))$.

Thus, in  the rest of this  paper, we will assume  that the state
spaces of the probabilistic models are compact convex subsets of a
locally compact topological vector space.  Notice that  this
assumption includes quantum theories (standard, statistical  and
relativistic)  and classical  theories as  well,  as particular
cases. We denote by $\C$ the set of states of a given probabilistic
model. The physical  interpretation of the convexity assumption is
that, given two states of the system, we should always be able to
form a convex combination of them, representing a  statistical
mixture. Convex sets    play   a    key   role    in   the formal
structure    of  quantum
theory~\cite{Holik-Ciancaglini,Holik-Zuberman}. The approach to
quantum theories   based   in   convex sets dates back  to the works
of B. Mielnik~\cite{Mielnik}  and G. Ludwig~\cite{Ludwig} (at
least). Recently,  the operational  approach based in convex sets
has attracted  much attention, related  to  the search of
operational and informational  axioms characterizing quantum theory
(see for
example~\cite{Entropy-PaperGeneralized,Entropy-generalized-II} and
references therein).

The extreme points of the state space are termed \emph{pure} states,
while other states are known as \emph{mixed} ones. As is  well
known, for the case of an  arbitrary (compact) convex   set  of
states $\C$ in finite  dimensions, each state $\nu\in\C$ can  be
written  as a convex  combination  of  its  extreme points. This is
indeed  the case in finite dimensional quantum  and classical
models~\cite{Entropy-PaperGeneralized}. In other  words, for each
state $\nu$, there exist a finite collection of extreme states
$\{\nu_i\}^{n}_{1}$, such that $\nu$ can be written as
\begin{equation} \label{e:Decomposition}
\nu = \sum^{n}_{i=1} p_i \nu_i.
\end{equation}
\noindent where $p_{i}\geq 0$ and $\sum^{n}_{i=1}p_i=1$. The state
space of a (finite-dimensional) classical model will be a
$d$-dimensional simplex, which can be defined as the convex hull of
$d+1$ linearly-independent points (defining a $d$-dimensional
simplex). In such a simplex, a point can be expressed as a unique
convex combination of its extreme points. It is remarkable that, for
Abelian C$^\ast$-algebras the state space is a simplex
(see~\cite[Vol.~1, Chap.~4]{Bratteli} and \cite[Chap.~10]{Phelps}
for more discussion on uniqueness of representing measures). Thus,
the decomposition in terms of extreme points will also be unique.
This characteristic feature of classical (commutative) theories no
longer holds in quantum models. Indeed, even in the case of standard
quantum mechanics  of finite dimensional  models, there are infinite
ways to express a mixed state as a convex combination of extreme
states.

In a  more general theory described  by a compact convex set $\C$,
the decomposition of a given state in terms of the set $\E(\C)$ of
extreme points of $\C$ is more involved (see~\cite{Bratteli,Alfsen}
for details). Given $\omega\in\C$ the goal is to build a
decomposition of the form
\begin{equation}
    \omega(a) = \int d\mu(\omega')\omega'(a)
\end{equation}
\noindent where $\mu$ is a measure over $\C$ supported by  the
extremal points of $\C$ and $\omega$ is  considered as  a
functional. This theory is related to the  theory of barycentric
decompositions in compact convex sets:  given  a  normalized Radon
measure  in  $\C$, its associated barycenter $b(\mu)$ will be given
by
\begin{equation}
b(\mu)=\int d\mu(\omega)\omega
\end{equation}
Given  a C$^\ast$-algebra $\M$ and  a (weak$^\ast$) compact convex
subset $\S \subseteq \C(\M)$,  it turns out that  for every state
$\omega \in  \S$, there exists a maximal\footnote{An order "$\leq$"
is introduced for the measures in $M_{+}(\C)$ as follows:
$\mu\leq\nu$, if and only if, $\mu(f)\leq\nu(f)$ for all real
continuous convex functions. A measure $\mu$ is said to be
\emph{maximal} with respect to "$\leq$" if, for all $\nu$ satisfying
$\nu\geq\mu$, we have $\nu=\mu$ \cite[Vol.~1, Chap.~4]{Bratteli}}
measure $\mu$, \textit{pseudosupported}\footnote{Given a compact
convex set $\C$, a measure $\mu$ is pseudosupported by the set of
its extreme points $\E(\C)$, if for each Baire set $B\subseteq \C$
satisfying $B\cap\E(\C)=\emptyset$, we have $\mu(B)=0$ \cite[Vol.~1,
Chap.~4]{Bratteli}.
 } in $\E(\C(\M))$~\cite{Ohya-1989,Bratteli}, such that
\begin{equation}
\omega=\int d\mu(\omega')\omega'
\end{equation}
The above result is much more general: it is valid for arbitrary
compact convex subsets of locally convex spaces (c.f.
\cite[Chap.~4]{Phelps} and \cite[Vol.~1, Chap.~4]{Bratteli}). As
usual in noncommutative models, the above decomposition is not
unique. For a given state $\omega$, we denote by $M_\omega(\C)$ the
set of all such measures.

\section{$(h,\phi)$-entropies}\label{s:Review}

In this section we discuss entropic measures in the  context of
standard quantum mechanics (i.e., we restrict our study to the case
of Type I factors) and return to the general setting in
Section~\ref{s:GeneralizedEntropies}. The $(h,\phi)$-entropies were
introduced by Salicr\'u \textit{et al.}~\cite{SalMen93} as follows:

\begin{definition}  \label{def:Salicru}  Let   us  consider
an $N$-dimensional probability vector  \ $p=[p_1  \: \cdots \:
p_{N}] \in  [0,1]^N$ with $\sum_{i=1}^{N} p_i = 1$. The so-called
$(h,\phi)$-entropies are defined as
\begin{equation} \label{eq:SalicruEnt}
\Salicru(p) = h\left( \sum_{i=1}^{N} \phi(p_i) \right),
\end{equation}
where  the \textit{entropic  functionals} $h:  \Rset \mapsto  \Rset$
and $\phi: [0,1] \mapsto \Rset$  are such that either: (i) $h$ is
increasing and $\phi$ is concave,  or (ii) $h$  is decreasing  and
$\phi$  is convex.  In both  cases, we restrict $\phi$  to be
strictly concave/convex and  $h$ to be  strictly monotone, together
with $\phi(0) = 0$ and $h(\phi(1)) = 0$.
\end{definition}

\noindent The family of $(h,\phi)$-entropies~\eqref{eq:SalicruEnt}
includes, as particular cases, the Shannon~\cite{Sha48},
R\'enyi~\cite{Ren61},
Havrda--Charv\'at--Tsallis~\cite{HavCha67,Dar70,Tsa88},
unified~\cite{Rat91} and Kaniadakis~\cite{Kan02} entropies. Dealing
with  the  infinite  dimensional context,  the  above definition
naturally  extends where the sum is  then over $i  \in \Nset$
provided the sum is finite (otherwise, by convention, the entropy is
set to be infinite).

In reference~\cite{QINP-paper}, a quantum mechanical version   of
the $(h,\phi)$-entropies was introduced and some  of their general
properties were discussed. But this definition  was restricted  to
finite dimensional quantum models. In  what follows  we advance a
definition for the  infinite dimensional case. We need to introduce
first the following concepts. Let us denote the set of
Hilbert--Schmidt operators acting on $\H$ by $\B_{HS}:= \{T \in
\B(\H): \Tr\left( T^2 \right) < \infty \}$ \cite{Holik-Zuberman}. As
is well known, the set $\B_{HS}$ endowed with the inner  product
$\braket{T_1,T_2} = \Tr\left( T_2^\dag T_1 \right)$ is a Hilbert
space. For $T \in \B(\H)$ the absolute value of $T$ is defined by
$|T|= \left( T^\dag T \right)^{\frac12}$. The subspace formed by the
trace class operators is defined as $\B_1(\H) = \left\{ T \in\B(\H):
|T|^{\frac12} \in \B_{HS}\right\}$. \emph{Quantum states} can be
defined as positive trace class operators of trace one (also called
\emph{density operators}). Now we can define the quantum
$(h,\phi)$-entropies.

\begin{definition} \label{def:QuantumSalicru}  Let us
consider  a quantum system described by a density operator $\rho$
(i.e., a positive trace class operator of trace one) acting  on  a
Hilbert space  $\H$. The quantum $(h,\phi)$-entropies are defined as
\begin{equation} \label{eq:QuantumSalicru}
\SalicruQ(\rho) = h\left( \Tr \phi(\rho) \right) ,
\end{equation}
where the \textit{entropic  functionals} \  $h: \Rset \mapsto \Rset$
\  and \ $\phi:[0,1] \mapsto  \Rset$ \ are such  that either:  \ (i)
$h$  is strictly increasing and  $\phi$ is strictly concave, or (ii)
$h$ is strictly decreasing and $\phi$ is strictly convex. We impose
$\phi(0)=0$ and $h(\phi(1))=0$ and we  take the convention
$\SalicruQ(\rho)=+\infty$  whenever $\sum_{i\in\Nset}\phi(p_i)$ is
not convergent. Here $\{ p_i \}_{i\in\Nset}$ is the sequence of
eigenvalues of the spectral decomposition of $\rho$, sorted in
decreasing order and counted with their respective multiplicities.
\end{definition}


\noindent The last convention  of this definition is justified as
follows. Every positive trace class operator $\rho$ of trace one
admits a spectral decomposition of the form $\sum_{i \in
\Nset}\lambda_i \mathbf{P}_i$, where $\{ \mathbf{P}_i\}_{i\in\Nset}$
is  a family of projection operators\footnote{Notice  that the
spectral decomposition  can  be easily  rewritten in  terms  of rank
one projections as $\rho = \sum_{i \in \Nset}s_i
\ketbra{\phi_i}{\phi_i}$, with $\sum_{i \in \Nset} s_i = 1$, $s_i
\geq 0$ and $s_i \geq s_{i+1}$. This  is known as   the
\emph{Schatten  decomposition}  of  $\rho$ (see,
e.g.,~\cite{Ohya-1989}).}. Thus, $\phi(\rho)
=\sum_{i\in\Nset}\phi(p_i)\mathbf{P}_i$. But then, $\SalicruQ(\rho)$
exists, only if $h\left( \Tr(\phi(\rho)) \right) = h\left(\sum_{i
\in \Nset}\phi(p_i) \right)<\infty$.

Regarding convergence in Definition~\ref{def:QuantumSalicru},  the
following remarks are  in order. Notice that $h\left( \sum_{i \in
\Nset}\phi(p_i) \right)$ will be  a convergent quantity --for all
$h$  and $\phi$-- whenever the rank of $\rho$ is  finite
dimensional. In principle, even if infinite dimensional ranks are
allowed, one may try to determine, given a particular choice  of $h$
and $\phi$, the set of states $\rho$  for which the sum  converges.
Notice also that for important families of entropic functionals,
$h\left( \sum_{i} \phi(p_i)\right)$ will be convergent for all
states.  As an example, consider the case  of the R\'{e}nyi
entropies with entropic index greater than one. A detailed study of
the  convergence properties of the infinite dimensional
$(h,\phi)$-entropies will be carried out elsewhere.

For finite dimensional $\H$, the Definition \ref{def:QuantumSalicru}
reduces to the one introduced in~\cite{QINP-paper}. Furthermore, the
quantum $(h,\phi)$-entropy  of a density operator $\rho$ equals the
classical entropy of  the sequence $p=\{ p_i \}_{i \in \Nset}$
formed by its eigenvalues: $\SalicruQ(\rho)=\Salicru(p)$.

An important notion for the rest of this work is that of
\emph{majorization}. Given two sequences $p=\{p_{i}\}_{i\in\Nset}$
and $q=\{q_{i}\}_{i\in\Nset}$ of positive real numbers sorted in
decreasing order, we say that $q$ is majorized by $p$ (and we denote
it by $q\preceq p$), if and only if,
$\sum_{i=1}^{k}q_{i}\leq\sum_{i=1}^{k}p_{i}$ for all $k\in\Nset$ and
$\sum_{i\in\Nset}q_{i}=\sum_{i\in\Nset}p_{i}$.

%


In what follows we make use of the integral form of the Jensen's
inequality (c.f. \cite{Niculescu}). Let $\mu$ be the Lebesgue
measure, $f:[a,b]\longrightarrow\Rset$ be a Lebesgue-integrable
function and $\phi$ a convex function. Then, for this case, Jensen's
inequality reads

\begin{equation}
\phi\left (\frac{1}{b-a}\int_{a}^{b} f(x)dx\right )\leq
\frac{1}{b-a}\int_{a}^{b} \phi(f(x))dx
\end{equation}

\noindent Assume that, for two sequences $p$ and $q$, we have that
$p\succeq q$. Thus, $q=Qp$ with $Q_{ij}=|U_{ij}|^{2}$ for some
unitary operator $U$ (see \cite{Li-Busch}). Due to the fact that
$\sum_{j\in\Nset}Q_{ij}=1$, for each $i\in\Nset$, we can decompose
the unit interval as

\begin{equation}
[0,1]=\bigcup_{k=0}^{\infty}\left[\sum_{j=1}^{k}Q_{ij},\sum_{j=1}^{k+1}Q_{ij}\right]
\end{equation}

\noindent (where we adopt the convention $\sum_{j=1}^{0}Q_{ij}=0$).
Put in words: we write the unit interval as an infinite union of
segments whose lengths are given by the sequence
$\{Q_{ij}\}_{j\in\Nset}$. Define a step function
$f:[0,1]\longrightarrow\Rset$ as $f(x)=p_{k+1}$ when
$x\in\left[\sum_{j=1}^{k}Q_{ij},\sum_{j=1}^{k+1}Q_{ij}\right)$ and
$f(1)=0$. By construction, we have that $\int_{0}^{1}f(x)dx=\sum_{k
\in \Nset} Q_{ik} p_k=q_{i}$ and $\int_{0}^{1}\phi(f(x))dx=\sum_{k
\in \Nset} Q_{ik}\phi(p_k)$. Thus, applying Jensen's inequality, we
obtain

\begin{equation}\label{eq:BistochasticConvex}
\phi(q_i)\leq\sum_{k \in \Nset} Q_{ik}\phi(p_k)
\end{equation}

\noindent Summing over $i\in\Nset$, we have

\begin{equation}
\sum_{i\in\Nset}\phi(q_i)\leq\sum_{i\in\Nset}\sum_{k \in
\Nset}Q_{ik}\phi(p_k)=\sum_{k \in \Nset}\phi(p_k)
\end{equation}

Let us now invoke  Theorem~4.1 of
Ref.~\cite{Arveson-Kadison}, that we reproduce here for
the sake of completeness. Let $\A$ be the  maximal
Abelian von Neumann generated  by   the  orthogonal  set
of  rank-one  projection   operators  $\{
\ketbra{e_k}{e_k} \}_{k  \in \Nset}$ and define the
conditional expectation map $E: \B(\H) \longrightarrow
\A$
\begin{equation}
E(A) = \sum_{n \in \Nset}\braket{ e_k | A |e_k}
\ketbra{e_k}{e_k}
\end{equation}
Given $\{  \lambda_n \}_{n \in \Nset}$,  a decreasing
sequence  in $\ell^1$ with non-negative  terms,  let
$\O_\lambda$  be  the  set  of trace  class  operators
possessing    $\lambda$    as    eigenvalue    list.
Then,    Theorem    4.1 of~\cite{Arveson-Kadison} asserts
that  $E(\O_\lambda)$ consists of all positive
trace-class operators $B \in \A$ whose eigenvalue list
$\{ p_n \}_{n \in \Nset}$ (arranged  in decreasing order)
is  majorized by  $\lambda$.  Then,  it follows that,  if
the  density operator  $\rho$  has an eigenvalue list
$\lambda =  \{ \lambda_n \}_{n  \in \Nset}$, the list
formed by $p  = \{ \braket{e_n|\rho|e_n} \}_{n \in
\Nset}$ (sorted  in decreasing order) is majorized by $\{
\lambda_n \}_{n \in  \Nset}$. Thus, as we have seen
above, for a convex function $\phi$ we have
$\sum_{n\in\Nset}\phi(\braket{e_n|\rho|e_n})\leq\sum_{n\in\Nset}\phi(\lambda_{n})$.
Remembering our convention for convex functions in
Definition \ref{def:QuantumSalicru}, we have

\begin{equation}
h\left(\sum_{n\in\Nset}\phi(\lambda_{n})\right)\leq
h\left(\sum_{n\in\Nset}\phi(\braket{e_n|\rho|e_n})\right).
\end{equation}

\noindent In other words, we obtain

\begin{equation} \label{eq:staticalmixtureC}
\SalicruQ(\rho) \leq \Salicru(p).
\end{equation}

\noindent A similar conclusion holds for the case of $h$ strictly
increasing and $\phi$ strictly concave. It is interesting to compare
inequality \ref{eq:staticalmixtureC} with Proposition $5$
of~\cite{QINP-paper}.

Due to  the fact that the  trace is invariant under arbitrary
isometries (i.e., transformations implemented by   operations
satisfying   $U^\dag U=\textbf{I}$\footnote{Notice that  all unitary
operators are  isometries}), it is easy to check that:

\begin{proposition}  \label{prop:unitary} The  quantum  $(h,\phi)$-entropies are
invariant under any isometric transformation  $\rho \to U \rho
U^\dag$ where $U$ is an isometric operator:

\begin{equation} \label{eq:unitary}
\SalicruQ(U \rho U^\dag) = \SalicruQ(\rho).
\end{equation}

\end{proposition}

Several properties of the family of entropies from
Definition~\ref{def:QuantumSalicru} were studied for the
case of finite dimensional Hilbert spaces
in~\cite{QINP-paper}. The study of the properties in the
infinite dimensional case is left for future work.

\section{Entropies in generalized probabilistic models} \label{s:GeneralizedEntropies}

In this section, we introduce an extension of the
definition of classical and quantum Salicr\'u entropies
to a more general family of probabilistic theories.

\subsection{$(h,\phi)$-entropies in C$^\ast$-algebras}\label{s:Cstar}

Let  us first  define  our extension  to  C$^\ast$-algebra models.  We follow  a
strategy that is analogous to that of~\cite{Ohya-1989}.

\begin{definition}
Given a C$^\ast$-algebra $\M$, for every
$\omega\in\C(\M)$, let:

\begin{eqnarray}
&D_\omega(\C(\M)) := \Big\{\mu \in M_\omega(\C(\M)) \quad \Big| &\\[2mm] \nonumber
&\exists \, \{ \mu_k \}_{k \in \Nset} \subset \Rset^+ \:
\mbox{ and} \quad \phi_k \subset \E(\C(\M)) \: \mbox{
s.t.} \: \sum_{k \in \Nset} \mu_k = 1 \: \mbox{ and}
\quad \mu = \sum_{k  \in \Nset} \mu_k \delta(\phi_k)
\Big\}&
\end{eqnarray}
where $\delta(\phi)$  is the  Dirac measure centered  at point $\phi$.  Now, for
$\mu \in D_\omega$ let
\begin{equation}
H(\mu) = h\left( \sum_{k \in \Nset} \phi(\mu_k) \right)
\end{equation}
when the above sum converges  and $H(\mu)=+\infty$ otherwise. Then,
by imposing to  the functions   $h$   and  $\phi$   the   same
conditions of Definition~\ref{def:QuantumSalicru}, we define
\begin{equation}
\SalicruQ(\omega) = \left\{
\begin{array}{l}
\inf \left\{H(\mu) \: | \: \mu \in D_\omega \right\} \\[2mm]
+\infty \quad \mbox{if} \quad D_\omega = \emptyset\\[2mm]
+\infty \quad \mbox{if} \quad \forall \, \mu \in D_\omega, \:
H(\mu)=+\infty
\end{array}\right.
\end{equation}
\end{definition}
It is important to remark that all models of standard quantum
mechanics are Type I factors  (Type I$_n$ for finite  dimensional
models and  Type I$_{\infty}$ for infinite dimensional ones), which
are C$^\ast$-algebras. For many measures, the above  definition
collapses  into the  one  of standard  quantum mechanics  when
restricted to Type I  factors. Indeed, if $\M$ is a Type  I factor,
by Gleason's theorem~\cite{Gleason},  every state $\omega$ can  be
described  by  a positive trace class operator $\rho_\omega$ of
trace one.  For finite dimensional models, we reobtain the  quantum
$(h,\phi)$-entropies of~\cite{QINP-paper}. For infinite dimensional
Hilbert  spaces, as is well  known, the von Neumann  entropy has the
same minimization  property (see for  example~\cite{Ohya-1989}).
Furthermore, Abelian C$^\ast$-algebras  are  in correspondence  to
classical statistical models.   Thus, our definition  contains also
an important  class of classical models as particular cases.

\subsection{More general models}

We now  discuss how to define  the $(h,\phi)$-entropies
in  an arbitrary compact convex set  $\C$, understood as
the state-space of a  generalized probabilistic model. We
will combine the approach presented
in~\cite{Entropy-PaperGeneralized} with the strategy used
in Section~\ref{s:Cstar}. Given a probabilistic model
described by a compact convex set $\C$,  let $\omega \in
\C$  be a state.  Denote  by $M_1(\C)$ to the set of
normalized Radon measures~\cite{Bratteli,Alfsen}. If
$\omega$ is the barycenter of $\C$ with respect to
measure $\mu$, we denote this by $\omega = b(\mu) = \int
d\mu(\omega') \omega'$. Define
\begin{equation}
M_\omega(\C) = \left\{ \mu \: | \: \mu \in M_1(\C) \quad \mbox{and} \quad \omega
= b(\mu) \right\}
\end{equation}
Now, in  analogy to  the procedure of
Section~\ref{s:Review}, we build  the set $D_\omega(\C)$,
and proceed in a similar  way as before.

\begin{definition}
Given a statistical theory whose state space is
represented by a compact convex set $\mathcal{C}$, define

\begin{eqnarray}
&D_\omega(\C) := \Big\{\mu \in M_\omega(\C) \quad \Big| &\\[2mm] \nonumber
&\exists \, \{ \mu_k \}_{k \in \Nset} \subset \Rset^+ \:
\mbox{and} \quad \phi_k \subset \E(\C) \: \mbox{ s.t.} \:
\sum_{k \in \Nset} \mu_k = 1 \: \mbox{ and} \quad \mu =
\sum_{k  \in \Nset} \mu_k \delta(\phi_k) \Big\}&
\end{eqnarray}

\noindent For $\mu \in D_\omega$, define
\begin{equation}
H(\mu) = h\left( \sum_{k \in \Nset} \phi(\mu_k) \right)
\end{equation}
\noindent when the above sum converges  and $H(\mu) =+\infty$
otherwise. Then, by imposing the same conditions on the functions
$h$  and $\phi$ as in Definition~\ref{def:QuantumSalicru}, we define

\begin{equation}
\SalicruQ(\omega) = \left\{
\begin{array}{l}
\inf\left\{H(\mu)) \: | \: \mu \in D_\omega \right\}\\[2mm]
+\infty \quad \mbox{if} \quad D_\omega = \emptyset\\[2mm]
+\infty \quad \mbox{if} \quad \forall \, \mu \in D_\omega, \:
\H(\mu) = \infty
\end{array}\right.
\end{equation}
\end{definition}
\noindent In  this way,  we obtain  a formal  expression
for the $(h,\phi)$-entropies in generalized probabilistic
models.

It is important to notice that, given a state $\omega \in
\C$, the  set $D_\omega(\C)$ can  be used to define a
notion of majorization  in generalized probabilistic
models in a similar way as
in~\cite{Entropy-PaperGeneralized}.

\begin{definition}
Suppose that there exists a  discrete measure $\tilde{\mu}$ such
that, for all $\mu \in D_\omega(\C)$, if we put $\{ \tilde{\mu}_i
\}_{i \in \Nset}$ and $\{ \mu_i \}_{i  \in \Nset}$ in decreasing
order, we have that $\sum_{i=1}^k \mu_i \leq \sum_{i =1}^k
\tilde{\mu}_i$ for all $k$ and $\sum_{i\in\Nset}\mu_i
=\sum_{i\in\Nset}\tilde{\mu}_i$. Then, by construction, we have that
$\{ \tilde{\mu}_i \}_{i \in \Nset}$ majorizes $\{ \mu_i \}_{i \in
\Nset}$ (and we write $\{ \mu_i  \}_{i \in \Nset} \preceq\{
\tilde{\mu}_i \}_{i \in \Nset}$) for  all $\mu \in D_\omega(\C)$. In
that case, we say that $\tilde{\mu}$ is the majorant of
$D_\omega(\C)$, and we call $\{  \tilde{\mu}_i  \}_{i \in \Nset}$
the spectra of $\omega$. Thus, if $\tilde{\mu}$ is the  spectra of
$\omega$ and $\tilde{\nu}$  is the spectra of $\sigma$, and we have
$\{\tilde{\nu}_i \}_{i \in  \Nset}  \preceq  \{\tilde{\mu}_i  \}_{i
\in \Nset}$, we then say that $\sigma\preceq \omega$ (i.e.,
\textit{$\omega$ majorizes $\sigma$}).
\end{definition}

\section{Final comments}\label{s:Conclusions}

In this short  paper, we have advanced a  definition of
the $(h,\phi)$-entropies for general probabilistic
theories,  extending previous definitions by including
(possibly) infinite  dimensional models. These examples
include  those of unital C$^\ast$-algebras and more
general compact convex sets.  Associated to the above
definitions, a natural definition  of majorization for
generalized probabilistic models arises (generalizing the
definitions presented
in~\cite{Entropy-PaperGeneralized}). A thorough study of
the properties of these entropic measures is left for
future work.

\section*{Acknowledgements}

MP, FH, PWL, GMB and GB acknowledge CONICET, UNLP and UBA
(Argentina), and MP and PWL also acknowledge SECyT-UNC (Argentina)
for financial support. SZ is grateful to the University of
Grenoble-Alpes and CNRS (France). MP acknowledges an AUIP grant and
warm hospitality at Universidad de Granada (Spain).


\end{document}